\documentclass[12pt,preprint]{aastex}
\usepackage{amsmath}
\usepackage{epstopdf}			
\newcommand{\etal}{\mbox{\rm et al.}}

\newcommand{\msun}{\mbox{$M_{\odot}$}}

\newcommand{\vsini}{\mbox{$v \sin i$}}

\newcommand{\fwhm}{\mbox{\rm FWHM}}

\newcommand{\rhk}{\mbox{$\log R^\prime_{\rm HK}$}}

\newcommand{\shk}{\mbox{$S_{\rm HK}$}}
\newcommand{\sten}{\mbox{\rm $S_{BL}$}}
\newcommand{\dels}{\mbox{\rm $\Delta S$}}
\newcommand{\sh}{\mbox{$S_{\rm H}$}}
\newcommand{\prot}{\mbox{$P_{\rm rot}$}}

\newcommand{\ms}{\mbox{m s$^{-1}$}}

\newcommand{\rms}{\mbox{rms}}
\newcommand{\snr}{\mbox{\rm signal-to-noise}}
\newcommand{\caii}{\ion{Ca}{2}}

\received{}
\accepted{}
  
\slugcomment{to appear in ApJ}
  
\shortauthors{Isaacson \& Fischer}
\shorttitle{Activity}
\begin{document}
  
\title{Chromospheric Activity and Jitter Measurements for 2630 Stars on the California Planet Search\altaffilmark{1}}
\author{Howard Isaacson\altaffilmark{2, 3}
Debra Fischer\altaffilmark{2, 4}}
  
\email{hisaacson@berkeley.edu, debra.fischer@yale.edu}
  
\altaffiltext{1}{Based on observations obtained at the Keck Observatory 
and Lick Observatory, which are operated by the University of California}
  
\altaffiltext{2}{Department of Physics \& Astronomy, 
San Francisco State University,
San Francisco, CA  94132}

\altaffiltext{3}{Department of Astronomy, 
UC Berkeley,
Berkeley, CA  94720}

\altaffiltext{4}{Department of Astronomy, 
Yale University,
New Haven, CT 06520}

\begin{abstract}
We present time series measurements of chromospheric activity for more than 2600 main sequence 
and subgiant stars on the California Planet Search (CPS) program with
spectral types ranging from about F5V to M4V for main sequence stars and 
from G0IV to about K5IV for subgiants. The large data set of more than 44,000 spectra 
allows us to identify an empirical baseline floor for chromospheric activity as a function of color 
and height above the main sequence. We define $\Delta S$ as an excess in emission 
in the Ca II H\&K lines above the baseline activity floor and define radial velocity 
jitter as a function of $\Delta S$ and \bv\ for main sequence and subgiant stars.
Although the jitter for any individual star can always exceed the baseline level, 
we find that K dwarfs have the lowest level of jitter. The lack of correlation 
between observed jitter and chromospheric activity in K dwarfs suggests that the observed  
jitter is dominated by instrumental or analysis errors and not astrophysical noise sources.  
Thus, given the long-term precision for the CPS program, radial velocities are not 
correlated with astrophysical noise for chromospherically quiet K dwarf stars,
making these stars particularly well-suited for the highest precision Doppler surveys. Chromospherically 
quiet F and G dwarfs and subgiants exhibit higher baseline levels of astrophysical jitter than 
K dwarfs. Despite the fact that the \rms\ in Doppler velocities is correlated with the mean 
chromospheric activity, it is rare to see one-to-one correlations between the individual 
time series activity and Doppler measurements, diminishing the prospects 
for correcting activity-induced velocity variations. 

\end{abstract}


\section{Introduction}
During times of high solar activity, the flux in an extended network of calcium lines brightens 
significantly in narrow band filters centered on the near-UV \caii\ lines.  
This observation led O. C. Wilson to anticipate that measurements of 
emission in the cores of \caii\ lines would provide a good index of stellar 
chromospheric activity and he began a survey at Mt. Wilson Observatory to 
search for analogs to the solar sunspot cycle \citep{wi68} in
bright nearby stars. 
Wilson developed a protocol for measuring \shk\ values, defined as 
the ratio of flux in the \caii\ line cores to flux in nearby continuum 
regions. These \shk\ values are now well established
indices for estimating chromospheric activity.  
\citet{du91} published seasonal \shk\ for 1296 stars 
observed as part of the Mt. Wilson Observatory (MWO) project from 1966 to 1983 and 
\citet{ba95} monitored a subset of 111 MWO stars to search for Maunder Minimum cycles. 
Other large scale surveys have also been carried out to monitor variability in chromospheric 
activity.
\citet{h96} derived \shk\ values for more than 800 nearby stars from the southern hemisphere;
\citet{st00} monitored \shk\ as a tracer for rotation period 
to find candidates for Doppler imaging; \citet{ha07} have measured H and K fluxes 
since 1994 to search for stars with sunlike spot cycles.  These surveys calibrate 
coefficients for the continuum and H \& K line cores so that the \shk\ values 
are on a scale that matches the original MWO \shk\ project. 

To account for different continuum flux levels near the \caii\ lines for stars of different 
spectral type, the \shk\ values are often parameterized as \rhk, a logarithmic fraction 
of the flux in the H and K line cores to photometric contributions from the 
star \citep{n84, m82}.  Because chromospheric activity 
declines with stellar age, cluster stars have been used to 
calibrate the \rhk\ values to the rotation periods and ages for 
stars with $0.4 < \bv\ < 1.0$ \citep{n84, mh08}.

Stars in the field rotate more slowly and have less chromospheric activity 
than their young cluster counterparts \citep{w63,k67,s72,s83, shb85,sd89}. 
\citet{gn85} find that most subgiants redward of G0IV also rotate 
slowly and have low chromospheric activity. \citet{sd89} measured UV emission 
as an activity indicator for a set of stars with masses between 1.2 and 1.6 \msun. 
They find that UV activity levels are essentially constant as lower mass stars with 
$M < 1.25 \msun$ evolve from the main sequence across the subgiant branch.
In contrast, stars with masses in the range of 1.25 - 1.5 \msun were found to 
exhibit moderate levels of activity (relative to the lower mass subgiants) 
on the subgiant branch with an abrupt drop in UV emission (activity) near 
spectral type G0IV that was empirically accompanied by a decrease in stellar 
rotational velocities. \citet{sd89} interpret this activity-rotation boundary 
as a physical transition from acoustic heating in early F-type main sequence 
stars to magnetic dynamo-driven activity followed by strong rotational braking
and a decrement in chromospheric activity.
As a case in point, \citet{g86} notes unusual high \vsini\ and 
moderate chromospheric activity in the F8IV star, HD~125840 and interprets this 
as a stage in subgiant evolution that might develop with the onset of intense 
dynamo activity as the convective zone deepens. 

Chromospheric activity is correlated with spots and flows in stellar 
atmospheres. High chromospheric activity is empirically associated with both random 
and quasi-periodic velocities that occasionally mimic the reflex velocities 
from exoplanets \citep{h02, q01, b07}. Therefore, it is a key parameter for 
identifying stars that may have photospheric features that produce radial
velocity ``jitter.''  Chromospheric activity 
is monitored in the stellar spectra used to derive precision radial velocities for 
stars on many Doppler planet searches \citep{sbm98,sf00,pch02,wr04,wr05,jfm07}.  
\citet{wr05} measured \shk\ for the California Planet Search (CPS) observations
obtained before 2004 August with the original HIRES CCD at Keck Observatory. 

Here we provide updated activity measurements with a calibration of \shk\ 
to the MWO scale for CPS spectra of main sequence and 
subgiant stars obtained after the upgrade of the HIRES CCD 
in 2004 August and activity measurements for stars on the Lick planet search project. 
We define an excess in \shk\ above a baseline floor (defined empirically as the tenth percentile 
level in chromospheric activity) as \dels\ for stars for the color range 
$0.4 < \bv\ < 1.6$ and examine correlations between \dels\ and radial velocity jitter. 

\section{Keck Observations and Analysis}

The CCD detector on Keck HIRES \citep{v94} was replaced in August 2004 with three new detectors 
that span a broader wavelength range and have higher quantum efficiency (particularly 
at blue wavelengths) and improved charge transfer efficiency. The HIRES B5 decker is 
generally used for CPS program observations and provides a spectral 
resolution of about $R = 52,000$.  The spectral format 
of HIRES is configured by adjusting the echelle and cross-dispersor angles so that (for CPS 
program observations) the iodine spectrum falls on the middle of the three chips. 
At the beginning of every CPS observing run, the focus algorithm is used to recenter a set of
Thorium and Argon emission lines so that the lines always fall on the same CCD pixels. 
This setup procedure ensures that stellar absorption lines fall on nearly the same part 
of the detector (modulo barycentric and stellar velocity shifts). This provides more 
consistency in the line spread function, relevant for high precision Doppler analysis.  

The blue CCD detector spans 3650 - 4790 \AA\  and includes the \caii\ H \& K lines used to 
assess chromospheric activity. Program observations are made through the iodine absorption cell, 
however, the molecular iodine lines disappear blueward of 5000 \AA\  and do not contaminate the \caii\ H\&K lines. 
This makes it possible to serendipitously measure stellar activity with every Doppler velocity 
measurement.

\subsection{Keck \shk\ Calibration}

Following a method described in \citet{du91}, we measure flux in the core of the 
\caii\ H \& K lines relative to continuum bands. Figures 1, 2 and 3 show the echelle orders 
used to measure \shk\ from a spectrum of the bright G8V star $\tau$ Ceti. The core flux 
in the \caii\ lines is weighted using a triangle function with a \fwhm\ of 1.09 \AA\ centered 
at 3968.47 and 3933.66 \AA\ for the H and K lines, respectively. The triangular weighting 
function gives the greatest weight to the core of the \caii\ lines, which are formed higher 
in the chromosphere, and diminishing weight away from the line core. Flux is also measured 
in two continuum windows. The continuum window on the redward side of the H and K lines is 
labeled as the $R$ section and the continuum window blueward of the lines is defined 
as the $V$ section. The R and V continuum sections monitor the overall flux of the star 
and help account for variable observing conditions such as changing air mass. As described 
below, coefficients for flux measurements in the line core and continuum windows are 
free parameters that scale the Keck data to match \shk\ values from the Mt. Wilson H \& K 
project for stars in common to both projects. 

Because the wings of the \caii\ lines stretch across a significant fraction of the echelle orders, 
it is difficult to define the continuum near these lines. As a substitute 
for continuum normalization, we construct a template for each star using  
five high \snr\ spectra for each star. Each of the five stellar spectra are 
shifted to rest frame wavelengths by cross-correlation with the National
Solar Observatory (NSO) solar atlas. 
A median filter is applied pixel by pixel in the stack of five observations to 
smooth over cosmic rays producing a high quality template. The time series spectra 
are then compared to the template, enabling precise measurement of variation in the \caii\ line cores.

For each of the time series spectra (program spectra, obtained for Doppler measurements), 
the echelle orders with the \caii\ lines are cross-correlated and shifted to 
the template wavelengths. The program spectra are then scaled in intensity to approxiately match the 
continuum intensity of the template spectra near the H \& K lines. The program 
spectra generally have different continuum slopes than the template because the barycentric 
velocity will differentially shift each of the time series spectral lines by several pixels.  
The blaze function of the echelle produces a maximum in flux near the center of 
the echelle orders that declines toward the red and blue edges of each order. 
As a result, the intensity is differentially affected by the steep blaze function across 
the order.  To align the continuum, each wavelength-shifted program spectrum is divided by the 
template spectrum and the residual is fit with a second degree polynomial. When 
aligning the continuum, the cores of 
the \caii\ lines are masked out so that any real variation in the line core flux is not affected.
The program spectrum is then divided by the polynomial so that it is 
finally well-aligned with the template both in wavelength and continuum intensity. 

After this careful alignment with the template spectrum, small relative differences 
in the \caii\ line cores can be measured to derive time series \shk\ values. Because
these measurements of chromospheric activity are simultaneous with the Doppler measurements, 
they provide a valuable diagnostic for interpreting prospective dynamical velocities in our
Doppler planet survey. 
	 
The calibration of coefficients for the continuum and core flux values was established 
using 151 main sequence stars on the CPS Keck survey that also have MWO \shk\ values published in 
the \citet{du91} catalog. The calibrating sample was restricted to stars with $B-V < 1.5$ that 
sit within one $V$ magnitude of the main sequence. We initially fit four scaling coefficients 
(Eqn 1) using a Levenberg-Marquardt (LM) algorithm:  
	
\begin{equation} 
\label{}
\shk\ =   \frac{ (C_{1} H + C_{2} K )} { (C_{3}R + C_{4} V)} 
\end{equation}

This produced covarient coefficients that were extremely 
sensitive to the initial guesses, so we reduced the number of free coefficients 
by fixing the ratio of flux meaurements in the H and K bands. 
\citet{wi68} and \citet{sh74} both note that the flux in 
the H and K lines should be equal in intensity, and that deviations from this are 
instrumental. Examining our spectra we find an average ratio of the H line flux 
to the K line flux of 1.45. In other words, multiplying the K line flux by 1.45 
boosts the K line intensity so that it is comparable to the H line. Similarly, we found that 
flux in the redder $R$ continuum window was dramatically higher than flux in 
the blueward $V$ continuum windows. We adopted a fixed coefficient of 25 to adjust flux in
the $V$ continuum window. Fixing the coefficients for $K$ and $V$ breaks the covariance 
between the remaining coefficients and simplifies the form of our 
calibration to Eqn~(2). The two coefficients in Eqn~(2) were determined
with Levenberg Marquardt fitting: 

\begin{equation} 
\label{}
\shk\ = C_{1}  \frac{( H + 1.45 K )} { (R + 25 V)} + C_{2} 
\end{equation}
\begin{equation} 
\label{}
\shk\ = 32.510  \frac{( H + 1.45 K )} { (R + 25 V)} + 0.021
\end{equation}

In Figure 4 we plot Keck \shk\ values derived with Eqn (3) against standard 
Mt. Wilson \shk\ values from \citet{du91}. We find an \rms\ scatter of 11\% to 
this fit that is likely the result of both measurement uncertainty (from MWO as 
well as from the Keck data) and intrinsic chromospheric variability.
Chromospheric activity of stars can vary on time scales of days to years. 
An RMS scatter of 11\% means that the \shk\ values published here are 
within 11\% of the long term average \shk\ of the MWO calibration stars. 

To empirically assess our measurement errors, we examined the RMS scatter for 
the chromospherically quiet, old G8V star, $\tau$ Ceti.  This star has a large number of 
observations at both Keck and Lick Observatory and long term \shk\ monitoring reported by 
\citet{du91} and \citet{wr04}. The distribution of \shk\ values measured at Keck 
are shown in Figure 5. The full width half maximum (\fwhm) of this 
distribution is about 0.002, our empirical assessment of measurement error. 
Thus, the measured \shk\ at Keck for five years of data for $\tau$ Ceti is $0.167 \pm 0.002$. 

Program spectra that have a $\snr < 5$ in the continuum near the \caii\ lines exhibit 
dramatically increased scatter in the measured \shk\ values. We adopt a minimum \snr\ of 
5 as a threshold for \shk\ measurements. Fewer than 2\% of the stellar 
spectra analyzed here had \snr\ below this rejection threshold.

\section{Lick Observations and Analysis}
The Lick Observatory Planet Search began in 1987 with the Hamilton Spectrometer (Vogt 1987) 
and both the Shane 3-meter telescope and the 0.6-m Coude Auxiliary Telescope (CAT).  The CCD detector 
at Lick was changed in 2002 to a Lawrence Berkeley Laboratory high resistivity CCD and the spectral 
format was extended at that time to include the \caii\ lines for simultaneous monitoring of 
chromospheric activity. The current spectral format ranges from 3800 to 9000 \AA\ with a spectral 
resolution of $R = 55,000$ at 6000 \AA\ .  The quantum efficiency (QE) of the detector is about 80\% for 
wavelengths between 5000 - 6000 \AA\ where the iodine absorption lines are analyzed to measure 
Doppler shifts.  The QE drops to less than 50\% near the blue \caii\ lines.  The spectra 
from the smaller aperture telescopes at Lick typically have lower \snr\ than Keck, particularly 
near the \caii\ lines.  

\subsection{Lick \sh\ Calibration}
The chromospheric activity measurements at Lick Observatory only make 
use of the \caii\ H line (Figure 6) because of inadequate \snr\ near the 
echelle order containing the \caii\ K line. Therefore, only an \sh\ 
value is calculated for spectra obtained at Lick Observatory. We adopt the approach of \citet{wr04} 
and measure the ratio of the line core flux in the \caii\ H line relative to a single 
continuum window, $C$: $L = H / C$.

Similar to our analysis of Keck stars, a coadded, median-filtered, 
template is created for each of the stars 
on the Lick program.  One additional step is made to eliminate cosmic rays or pixels affected 
by Compton scattering of electrons in the CCD for program observations.  This is more of a problem at 
the Lick Observatory because of the longer exposure times and a thicker substrate in the 
high resistivity CCD, which is more subject to Compton scattering. 
To identify affected pixels, the wavelength-shifted and flux-scaled 
program spectrum is divided by the template observation.  Pixels with values that are 
ten sigma away from the median value are replaced and the program spectrum is then 
iteratively realigned with the template.  Spectra with replaced pixels in the continuum 
window or in the line core are flagged and excluded if they are outliers so that 
cosmic ray cleaning does not affect the \sh\ measurement; it simply improves the 
alignment of the program observations with the template. 

Coefficients for the Lick \sh\ values (Eqn~4) were calibrated using 83 stars in common with
\citet{du91}. Only stars with a color $B - V < 1.6$ that 
were within one $V$ magnitude of the main sequence were used in the calibration. 
A Levenberg Marquardt fitting analysis was used to find the coefficients that minimized 
the RMS scatter between the Lick \sh\ values and the \citet{du91} \shk\ measurements. 
This calibration has an \rms\ scatter of 11\%, shown in Figure 7.
We tested inclusion of higher order terms but did not find improvement in the 
\rms\ fit. 

\begin{equation}
\label{ }
\sh\ = 2.206  L^{2} + 6.907  L  
\end{equation}

To estimate our \sh\ measurement uncertainty at Lick Observatory, we again 
considered time series \sh\ measurements for the chromospherically inactive 
star, $\tau$ Ceti. A histogram of the 754 \sh\ measurements for this star is 
plotted in Figure 8. The \fwhm\ is 0.008 and represents 
the empirical uncertainty in our single measurement precision for \sh\ at Lick Observatory. 
This single measurment precision is about five times larger than the Keck \shk\ values, 
but the median \sh\ value from Lick is $0.1627 \pm 0.008$ and agrees with the Keck 
value \shk\ $= 0.167 \pm 0.002$ within uncertainties.  

\section{Results} 
In total, more than 44,000 spectra of 2630 stars were analyzed and calibrated 
to the Mt. Wilson \shk\ activity index.  The \shk\ and \bv\ values were used to 
calculate \rhk, a measure of the flux in the \caii\ lines relative to 
the basal photospheric value \citep{n84}. While \shk\ has a functional dependence on
spectral type, \rhk\ removes the basal component of core emission and 
only varies with chromospheric activity. The \rhk\ parameter has been calibrated to rotation periods
and stellar ages for FGK stars \citep{n84,mh08}. 

The main disadvantage of \rhk\ is that stars 
with \bv\ less than 0.4 or greater than 1.0 were not included in the original calibrations
because main sequence stars outside this range were not observable in
distant clusters \citep{n84,mh08}. The derived quantities \rhk, \prot, and 
stellar age are most secure for main sequence stars that have \rhk\ between -4.0 and -5.1 and 
\bv\ values between 0.4 and 1.0. 

Table \ref{T1} contains a 
summary of star names, \bv color, the median values of \shk\ and \rhk\, height ($\delta M_V$) 
above the main sequence, \prot\ (days), Age (Gyr) and the Observatory 
where the measurements were obtained (Keck or Lick). 
A stub of this table is provided in the printed version of this paper and the 
complete table is available in the online version of this paper. 

For evolved stars and stars outside the color range $0.4 < \bv < 1.0$, the derived parameters 
\rhk\, \prot\ and Age were not calibrated by \citet{n84} or \citet{mh08} and are not included here.
The complete time series data from Keck and Lick Observatories are listed in Table \ref{T2}. 
The spectra from Keck Observatory were collected between 2004 August and 2010 August. 
These values can be appended to results contained in \citet{wr04} so that for many stars, 
the measurements approach the span of an entire activity cycle (e.g., an eleven year solar cycle).  
The time series observations in Table \ref{T2} (online data only) from Lick Observatory span 
2002 through 2009 December. 

\subsection{Outliers}
A few specific stars have values very different from the published values of 
\citet{du91} and \citet{wr04}. 
In \citet{wr04}, the \shk\ value for HD~38392 has a typographical error \citep{wpvt09}, HD~104958 has 
an \shk\ of 0.07 which is likely a typographical error (because it is non-physical) and 
HD~220339 has an \shk\ value of 0.000, which is also non-physical \citet{wr04}. 
We also note that HD~195019 is a binary with S-values for the A and B components. 

\subsection{Chromospheric activity of main sequence stars} 
Stars in the CPS sample are defined here as main sequence stars if the absolute 
visual magnitude is within 1.5 magnitudes of the main sequence. 
The simultaneous measurements of activity and precision velocities
are an ideal data set to search for correlations between activity and spurious radial velocity 
jitter. Unfortunately, \rhk\ has not been calibrated for evolved or late type stars. An 
activity metric is desirable because these stars are being monitored for exoplanet surveys 
and we would like to understand the impact of chromospheric activity on radial velocities for 
evolved and late-type stars (as well as main sequence stars) to quickly assess whether 
velocity scatter is likely to be astrophysical or dynamical in origin.  

As a proxy for \rhk, which accounts for the basal (rotation independent) 
photospheric flux near the H \& K lines, we define baseline activity values, \sten, by 
fitting the floor in \shk\ for main sequence 
stars as a function of \bv\ color: 

\begin{align} 
\sten  =  & 2.7 - 16.19(B-V) + 36.22(B-V)^2 - 27.54(B-V)^3 -14.39(B-V)^4 \\
\nonumber &  + \ 34.97(B-V)^5 -18.71(B-V)^6 +3.17(B-V)^7 
\end{align}

The polynomial fit described by Eqn~5 for \sten\ is shown in Figure 9 as a dashed red line and was obtained by selecting 
the tenth percentile (i.e., inactive) \shk\ values in twelve \bv\ bins (0.1 magnitude in width). 
Statistical error bars were calculated by dividing the tenth percentile activity value by 
the square root of the number of points in each bin. The error bars were used to
weight the polynomial fit to the baseline \shk.  Because of the large stellar 
sample, we were able to carry out this empirical fit, rather than relying on theoretical 
estimates to remove the basal emission in the \caii\ line core. It is reasonable to assume 
that each of the color bins contains a fraction of low activity stars. 

The \sten\ values for main sequence stars begin to rise redward of $\bv = 1.0$, 
expected in part because of a decrease in continuum flux for redder stars.  Figure 10 demonstrates
the genuine range in \caii\ core emission, plotting \caii\ lines for three stars with $\bv \sim 1.5$
but different \shk\ values. The lower envelope for \shk\ turns over at \bv\ of 
1.4. This is consistent with activity measurements of \citet{rm06} who measured line 
core equivalent widths and show that the luminosity in the \caii\ lines decreases by a 
factor of $\sim3$ with decreasing mass from K7 to M5V. Likewise, 
\citet{grh02} measured H$\alpha$ EW as an indicator of chromospheric 
activity in M dwarfs and found that the EW of absorption lines decreases with 
increasing spectral type from K7V to M4V. Models of \citet{cg87} show that 
M dwarfs with the strongest H$\alpha$ absorption have moderately active chromospheres, 
while weak H$\alpha$ correlates with weak chromospheric activity. 
\citet{cg87} further show that activity in M dwarfs may be relatively constant
over time, in contrast with solar type stars with strong activity cycles. 

\subsection{Chromospheric activity in subgiants} 

The activity measurements in this paper include a substantial subset  
of 234 subgiants, included a survey that targets subgiants \citep{jfm07, j10} and contained in 
the N2K survey \citep{f05}. The \shk\ values for these stars, plotted in Figure 11,  
show a modest delcine in the median \shk\ value as a function of \bv. A typical \shk\ 
value for subgiants is $0.1 < \shk < 0.2$.  However, 10\% of the 234 stars show moderate 
to strong chromospheric activity, with \shk\ values greater than 0.3.  We have identified 
some of these active stars as spectroscopic binaries (blue dots).  \citet{gmd98} and \citet{grh02} have noted 
a correlation between chromospheric activity and orbital periods for both main sequence and evolved 
stars. In Figure 12, the median \shk\ value is plotted as a function of \vsini. Stars 
with higher \vsini\ are also more active, consistent with the idea proposed 
by \citet{g86} that activity may be a transient phase for subgiants. However, the 
sprinkling of active stars is found across the subgiant branch, not just at the G0IV boundary 
identified by \citet{sd89}.

\section{Activity and jitter}
The primary motivation for studying chromospheric emission in planet search 
stars is to monitor potential sources of astrophysical noise that can impact 
the radial velocities. We calculate the difference between the median \shk\ value and 
\sten\ as a \dels\ for each star. This \dels\ is an empirical parameterization for 
the excess activity of stars at all \bv\ colors, including late K and early M dwarfs. 
Because the photon-limited velocity precision at Keck is better than Lick, we restricted 
our assessment of velocity jitter to stars observed with HIRES at Keck. 

We divided the target stars into four \bv\ groups and plotted velocity \rms\ 
as a function of \dels\ in Figure 13. We did not remove any linear trends or 
periodic velocities, even from stars with known planets, so some of the scatter 
above the lower envelope of velocity \rms\ is due to dynamical velocities. We 
define jitter as the quadrature difference of velocity \rms\ minus the formal 
internal errors (Eqn~6). We emphasize that since velocity scatter includes 
instrumental and analysis errors, this definition of jitter includes systematic instrumental 
errors as well as astrophysical noise. Furthermore, the (astrophysical or 
instrumental) jitter for any particular star may be higher than the tenth percentile velocity \rms\ floor.

\begin{equation} 
\label{}
{\rm Jitter} = \sqrt{\rms^2 - \sigma^2}
\end{equation}

Figure 13 (top, left) contains data for 259 late F to mid G-type 
main sequence stars with $0.4 < \bv\ < 0.7$. 
The minimum jitter as a function of \dels\ is fit to the lower tenth percentile in velocity \rms\ 
with a line that is plotted in red in the top left panel of Figure 13. 
For chromospherically quiet stars, the level of jitter 
begins at 2.3 \ms and increases rapidly as a 
function of \dels\ (Eqn~7).  Note that the \dels\ range is smaller for the blue 
stars because the photospheres are brighter at blue wavelengths 
and there is less contrast in the \caii\ lines.  The stars on the CPS sample 
in this bluest \bv\ color bin exhibit weaker \caii\ core emission, 
however, the \rhk\ value for these stars climbs steeply as a function of increasing \shk. 
In our experience, these stars have the strongest velocity jitter of the four groups considered.  

The second group consists of 218 main sequence stars in the color range $0.7 < \bv\ < 1.0$. 
The tenth percentile floor in velocity \rms\ is fit by Eqn~8 and plotted
as a red line in the top right panel of Figure 13. The minimum jitter for chromospherically 
quiet stars in this \bv\ color range is 2.1 \ms\ and jitter increases with increasing
activity, \dels.

The third color bin, $1.0 < \bv\ < 1.3$ contains 118 stars. These stars have a minimum 
noise floor of 1.6 \ms\ and Eqn~9 does not have a dependence on \dels.  Even when the 
emission in the core of the \caii\ lines is significant, it is possible to find stars 
in this color bin that have low jitter. 

The noise floor for the 89 reddest stars (bottom right panel of Figure 13) begins at 2.14 \ms\ 
and Eqn~10 shows that jitter increases slowly with $\dels\ $. It is possible that the noise for this 
group of stars is affected by analysis errors, since the deconvolution is 
particularly challenging for spectra of late type stars. 

\begin{equation} 
\label{}
{\rm Jitter} = 2.3 + 17.4 * \dels\ \ms\ \,\, (0.4 < \bv\ < 0.7) 
\end{equation}
\begin{equation} 
\label{}
{\rm Jitter} = 2.1 + 4.7 * \dels\ \ms\ \,\, (0.7 < \bv\ < 1.0) 
\end{equation}
\begin{equation} 
\label{}
{\rm Jitter} = 1.6 - 0.003 * \dels\ \ms\ \,\, (1.0 < \bv\ < 1.3) 
\end{equation}
\begin{equation} 
\label{}
{\rm Jitter} = 2.1 + 2.7 * \dels\ \ms\ \,\, (1.3 < \bv\ < 1.6)
\end{equation}

We have also carried out an analysis of activity-correlated jitter for subgiant stars observed at 
Keck, defined here as stars that have a $\delta M_V$ of at least 1.5 magnitudes 
above the main sequence. Eqn~11 provides our fit to the bottom tenth percentile activity values.
Ninety percent of subgiants have low chromospheric activity, so the 
fit to the floor of the velocity \rms\ in Figure 14 spans a relatively 
narrow range of \dels. The jitter for subgiants is defined in Eqn~12.
\begin{equation} 
\label{}
\sten =  0.2 - 0.07 * \bv 
\end{equation}
\begin{equation}
\label{}
{\rm Jitter} = 4.2 + 3.8 * \dels\ \ms\ \,\, (\delta M_V > 1.5)
\end{equation}

\section{Discussion}
The \caii\ lines are good indicators of chromospheric activity for main 
sequence stars.  Here, we present measurements of this emission, parameterized as \shk\ values,
for $\sim 2600$ stars observed at Keck Observatory and Lick Observatory.  The time series 
of activity measurements at Keck Observatory date back to 2004 August and the time series 
data for Lick began in 2002 when the spectral format was extended to include the \caii\ lines. 

To put \shk\ on a continuous time baseline, we calibrated our measurements to the values 
obtained from the long-standing Mt Wilson Observatory H \& K program \citep{wi68,du91} using 
a set of stars common to both projects. Although we developed a new approach for 
differentially measuring \shk, these time series activity measurements may be 
appended to the \citet{wr04} values, which were also calibrated to the Mt Wilson H\&K program. 
We find an \rms\ of 11\% in the difference between our \shk\ values and those on 
the long term Mt Wilson H \& K project.  Since the \rms\ of \shk\ values for chromospherically 
quiet stars is only 1\% for Keck stars and 5\% for Lick stars, we expect that long term 
chromospheric variability contributes to the 11\% \rms\ in our calibration. 
Our measurement of \shk\ includes a significant sample of stars redward of $\bv\ = 1.0$.

The Mt Wilson H \& K project surveys the brighter F, G and K-type main sequence 
stars and subsequent calibrations to \rhk, \prot\ and ages were restricted to these
spectral types.  With the benefit of a large statistical sample of
stars, we define the lower envelope of activity, \sten, as a function of \bv\ for
main sequence and subgiant stars over the color range $0.4 < \bv\ < 1.6$.
We define \dels\ as the difference between the median \shk\ for each star and \sten.
We then evaluated velocity jitter for our stars 
by subtracting the mean internal error from the \rms\ of the velocities for each star. 

Velocity ``jitter'' therefore characterizes the extent to which the velocity \rms\ deviates 
from our $1 \sigma$ internal errors. Jitter includes instrumental and analysis errors as well as 
photon shot noise, unresolved dynamical velocity shifts, and photospheric noise, and it is not 
always clear which of these is the dominant term.  In programs with uniform analysis precision 
where a constant \snr\ is obtained for all observations (the case for CPS observations of bright 
stars at Keck), it is possible to treat jitter as a free parameter in the Keplerian model. However, both Frequentist and Bayesian models 
can benefit from physical priors since additional unresolved low amplitude planet signals 
will inflate residuals to the Keplerian fit. It is likely that jitter will be over-estimated (and more ambiguous) 
if velocities are obtained with a spectrometer where hardware components (e.g., the detector and controller) 
have changed, or if data sets have been obtained with more than one telescope, or if the \snr\ changes 
significantly from observation to observation. 

A key conclusion from this work is that K dwarfs in the color range $1.0 < \bv\ < 1.3$
have the lowest level of velocity jitter.  Importantly, jitter appears to be decoupled from chromospheric 
activity in these stars; it is a constant 1.6 \ms\ independent of the strength of emission in 
the \caii\ line cores. The lack of correlation between observed jitter and chromospheric 
activity in K dwarfs suggests that jitter for these stars is dominated by instrumental or analysis errors 
and not astrophysical noise sources. This demonstrates two important 
points: (1) on all timescales since 2004, the systematic instrumental errors in CPS observations at Keck 
are no more than 1.6 \ms\ and (2) the mid to late K dwarfs represent a sweet spot for optimal detectability in 
Doppler exoplanet searches.  The astrophysical contribution to velocity noise in K dwarfs
could be significantly less than 1 \ms.

Relative to Kdwarfs, we also show that jitter increases as a function of \bv\ for chromospherically inactive stars. 
The lowest activity stars in the bluer color bin of $0.7 < \bv\ < 1.0$ and in the redder color bin of 
$1.3 < \bv < 1.6$ exhibit a measurably higher minimum jitter of about 2.1 \ms\ and a weak dependence 
on activity (\dels). The minimum jitter increases slightly to 2.3 \ms\ for chromospherically 
inactive late F and early G dwarfs ($0.4 < \bv < 0.7$) and there is a strong sensitivity to \dels.

If we assume that most of the observed K dwarf jitter is instrumental in origin (i.e., we assume that the astrophysical 
contribution to jitter is essentially zero for K dwarfs), then stars in other color bins 
exhibit a quadrature sum of 1.6 \ms\ instrumental noise plus additional intrinsic astrophysical noise. 
Thus, excluding instrumental contributions, the astrophysical noise floor imposed by chromospherically inactive late F 
and early G dwarfs appears to be at least $\sqrt{ 2.3^2 - 1.6^2}$ = 1.7 \ms\ and we expect that it will 
be difficult, even with the best Doppler precision, to beat down this noise floor. 
It may be possible to reduce the impact of stellar jitter in late F and G-type dwarfs by confining Doppler searches to 
stars in the bottom quartile of activity level and by observing the stars very frequently (i.e., with a 
high cadence strategy). However, in general, solar twins are not optimal targets for Doppler exoplanet 
searches that aim to detect very low mass planets. At best, sunlike stars will certainly require more 
data points to average down a noise floor that starts out higher than the astrophysical noise from K dwarfs. 

By this same reasoning, mid- to late-G dwarfs with $0.7 < \bv\ < 1.0$ impose an astrophysical 
noise floor that is $\sim1.3$ \ms\ above that of K dwarfs. It is possible that later type stars with $1.3 < \bv\ < 1.6$ 
may have greater jitter because of challenges with deconvolution (i.e., a noise source that is not astrophysical). 
A dependence of \rms\ velocities on \dels\ would argue for astrophysical noise, however the functional dependence 
is quite weak and the number of stars with larger \dels\ puts us in the realm of small number statistics. 

Since there is a correlation between \dels\ and the \rms\ velocity floor in stars, one might hope 
that variations in individual observations of \shk\ would have a nearly one-to-one correlation with Doppler 
velocity measurements. \citet{q01} reported the remarkable case of HD~166435, a young G0V star with quasi-period 
radial velocities that were correlated to line bisector variations and \caii\ emission. In this unusual case, 
the star seemed to have persistant active longitudes where spots were regenerated for almost two years, producing 
a coherent signature of magnetic activity.  However, stars do not generally show such clear correlations between 
activity and velocity.

A more typical example is HD~143714, a metal-rich G0V star from the N2K program with $\bv\ = 0.61$ and a 
median $\shk\ = 0.24$. According to Eqn~5, $\sten\ = 0.147$ for a star of this color, so $\dels\ = 0.093$. 
Using Eqn~7, the expected jitter for HD~143714 is 3.9 \ms. By the end of 2008, the velocities for HD~143714
showed a significant \rms\ of almost 10 \ms\ and were periodic (Figure 16).  However, we observed an apparent 
correlation between the Doppler velocities and activity (Figure 17).  When the velocities were detrended
using the linear fit to activity, the periodicity disappeared and the \rms\ dropped to 6.2 \ms.  We apparently 
observed an epoch where chromospheric activity and radial velocities were correlated and the level of velocity 
\rms\ exceeded the predicted jitter level.  However, the activity correlation disappeared. Velocity measurements 
obtained after 2009 January are no longer correlated with activity and have an \rms\ of 4.9 \ms\, closer to the 
predicted jitter.  Similar correlations between \shk\ and velocities have been observed in only a few other CPS stars.
Like HD~166435 and HD~143174, the stars are late F or early G dwarfs and the correlations come and go. 
The distribution of jitter values for CPS stars is summarized in the HR diagram in Figure 15. 

Subgiants generally exhibit lower chromospheric activity than main sequence stars, 
however these stars apparently have a jitter floor starting at about 4 \ms, significantly 
higher than the jitter level of main sequence stars with similar \bv. 
As the stars expand, the depth of the convective zone increases, braking rotation and 
decreasing the convective turnover timescale. Consistent with this, ninety percent of 
the subgiants analyzed here have low chromospheric activity. However, ten percent of 
the subgiants have moderate to strong chromospheric activity that is correlated with 
rotational velocity (\vsini) or the presence of a close binary companion. These active 
subgiants are reminicent of the F8IV star, HD~125840. \citet{g86} explained the unusual 
activity of this star as the onset of a convective dynamo during the evolution of the 
massive progenitor star. Activity in subgiants has been observed by \citet{gn85} and 
\citet{sd89} for subgiants hotter than G0IV.  However, the active subgiants here 
are not limited to this boundary region; they are found across the subgiant branch. 
This range of subgiant activity may reflect a range of conditions in the progenitor 
star, such as stellar mass or rotation, that transitions through a convective dynamo 
stage at different points on the subgiant branch. Because only a small fraction of 
subgiants are active, this transition phase is likely short-lived. \citet{sd89} claim
that activity in subgiants is restricted to 100 Myr; the time for early F-type 
stars to evolve from the main sequence to a G0IV star, implying that the convective 
zones of most subgiants does not sustain convective dynamos for long. Although 
the phenomenon of activity in subgiants is uncommon, the existence of active subgiants
with spectral types later than G0IV suggests that this phase can persist longer than 
the 100 Myr estimate by \citet{sd89}.

\acknowledgements
We thank Jason T. Wright for useful conversations. 
We gratefully acknowledge years of work on the Keck planet search program by G. W. Marcy, 
R. P. Butler and S. Vogt. We also acknowlege the many observers who 
collected observations and contributed to the Doppler analysis: J. A. Johnson,
A. W. Howard, Julien Spronck, Kelsey Clubb, John Brewer, K.Peek, J. Anderson,  J. A. Valenti, N. Piskunov.  
We gratefully acknowledge the dedication and support of the Keck
Observatory staff, in particular Grant Hill and Scott Dahm for support with HIRES 
and Greg Wirth for supporting remote observations. 
D.A.F. acknowledges research support from NASA grant NNX08AF42G and NSF AST 1036283.
We thank the NASA Exoplanet Science Institute (NExScI) for 
support through the KPDA program. We thank the NASA and NOAO 
Telescope assignment committees for allocations of telescope time.  
The authors extend thanks to those of Hawaiian ancestry on whose 
sacred mountain of Mauna Kea we are privileged to be guests.  
Without their kind hospitality, the Keck observations presented 
here would not have been possible. This research has made use of 
the SIMBAD database, operated at CDS, Strasbourg, France, and of 
NASA's Astrophysics Data System Bibliographic Services.

\begin{figure}
\plotone{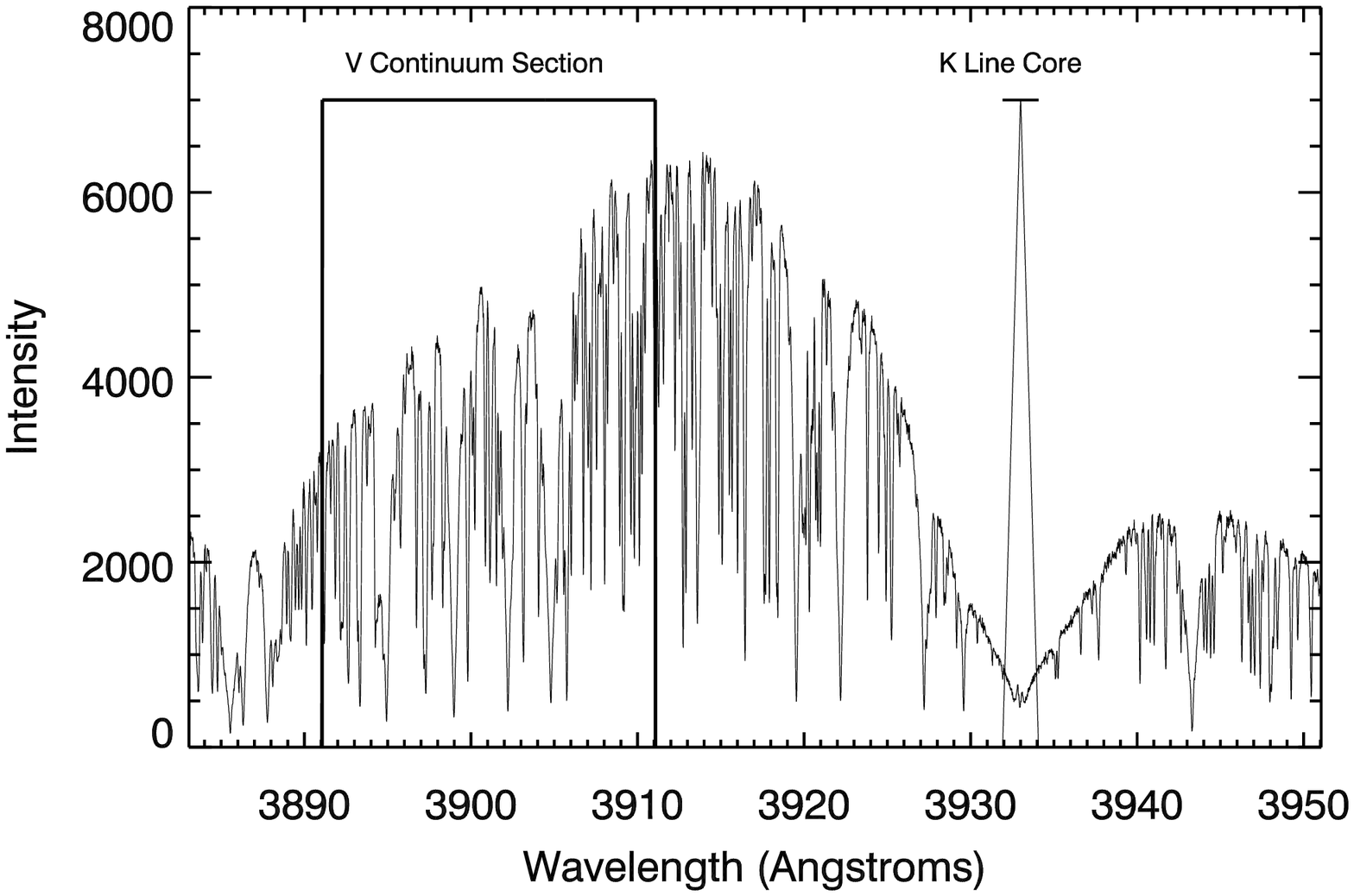}
\figcaption{The echelle order from Keck/HIRES containing the \caii\ K line for the 
G8V star, $\tau$ Ceti.  Emission in the 
core of the \caii\ K line is weighted by the indicated normalized triangle, giving greatest 
weight to flux in the core of the line. The \caii\ flux is measured relative to continuum 
flux in the indicated $V$ band.} 
\end{figure}
\clearpage

\begin{figure}
\plotone{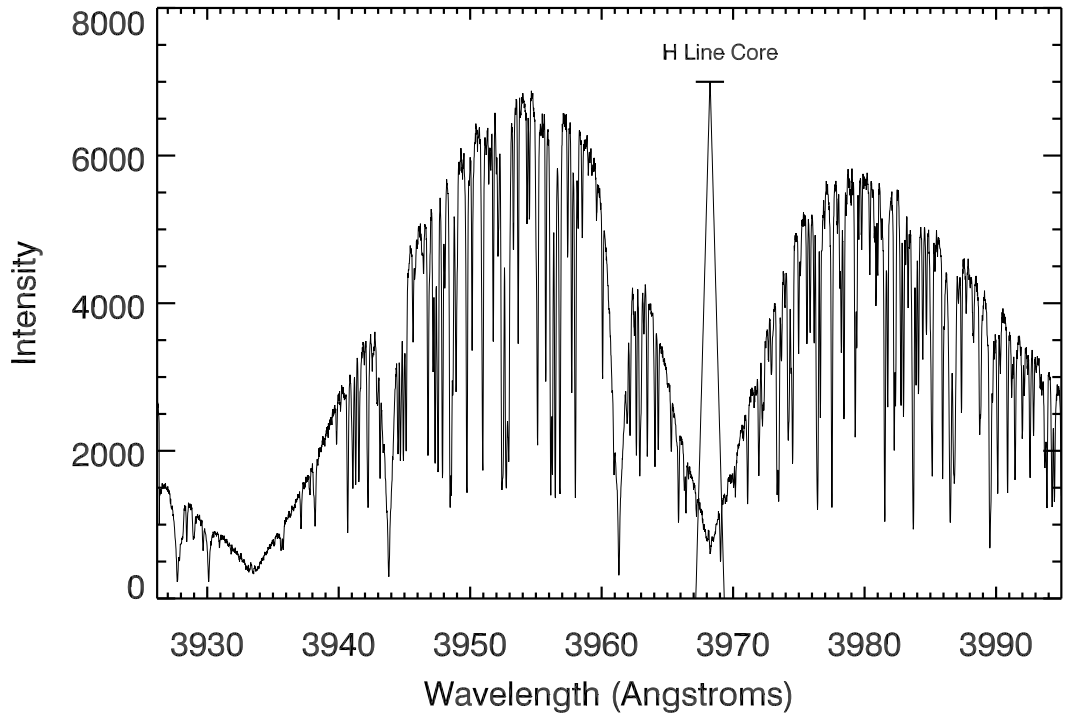}
\figcaption{The echelle order from Keck HIRES with the \caii\ H line near the center of the 
echelle order.  The K line toward the edge of the detector where the blaze function is dropping was 
not used because of the low SNR in the line and continuum flux.
The triangular weighting function is indicated on the plot for the H line.  The 
wings of the H \& K lines in this order suppress the continuum so that the $R$ continuum band is 
defined in an adjacent order. }
\end{figure}
\clearpage

\begin{figure}
\plotone{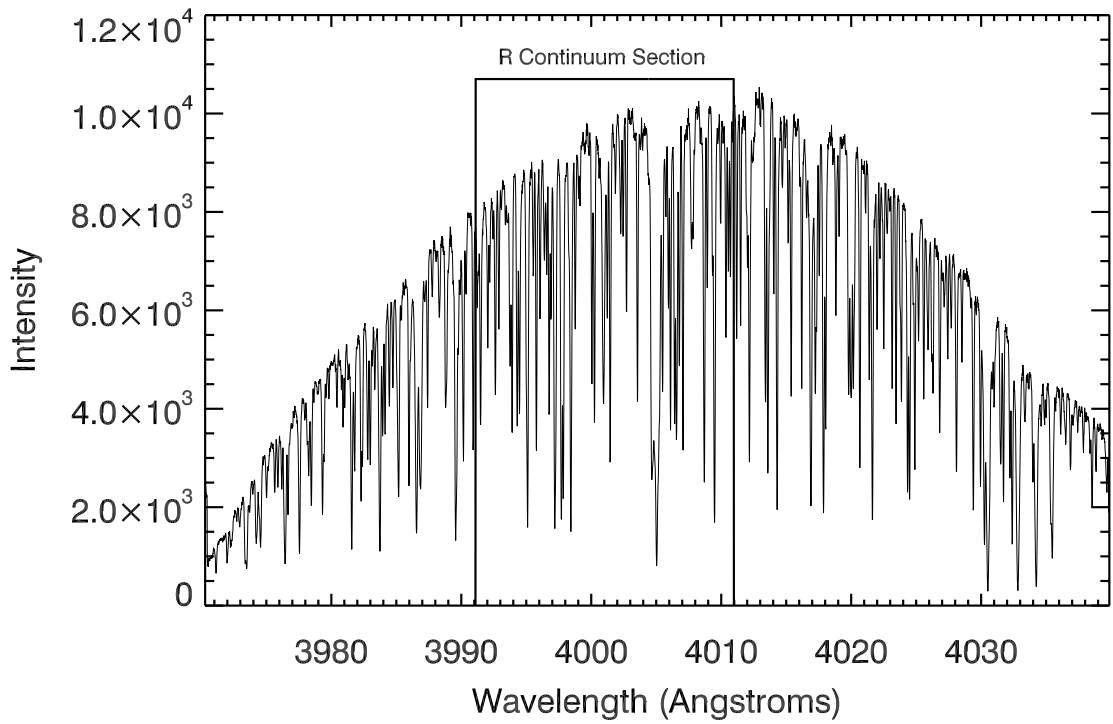}
\figcaption{The 20 \AA\ $R$ band was defined in an adjacent order to monitor photospheric 
continuum redward of the \caii\ H \& K lines. This spectrum in this figure is a 
Keck/HIRES spectrum of $\tau$ Ceti.  }
\end{figure}
\clearpage

\begin{figure}
\plotone{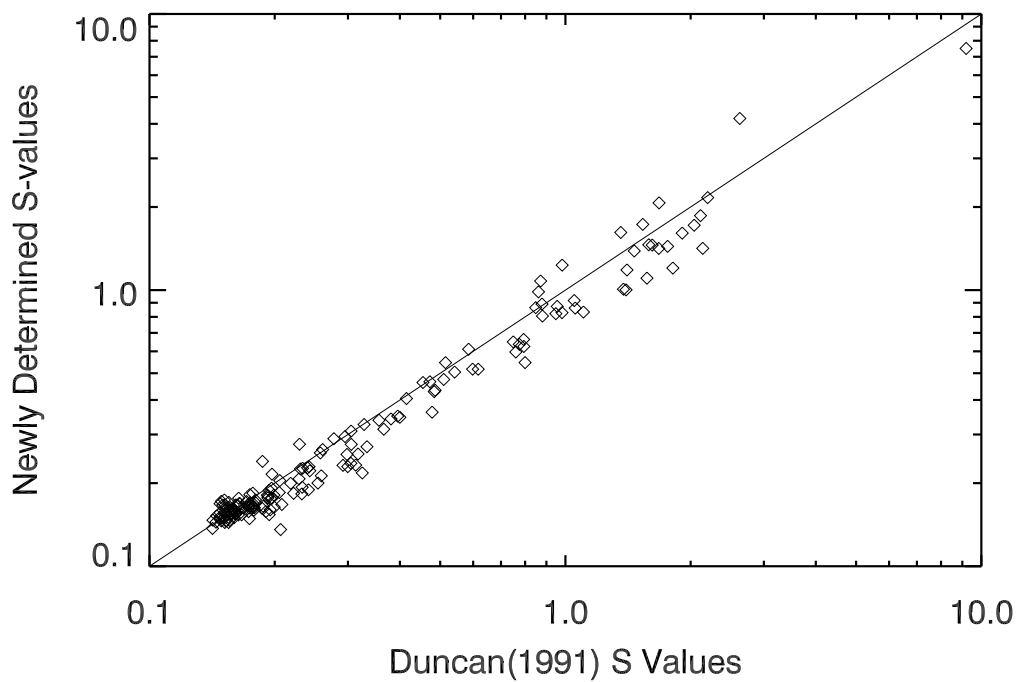}
\figcaption{The calibration of coefficients for \shk\ was made using 151 stars
observed in common with the Mt. Wilson project \citep{du91}. The free parameters (coefficients) 
were determined with a Levenberg Marquardt algorithm that minimized the difference 
between Keck and Mt. Wilson \shk\ values. The calibration yielded an RMS scatter of 11\% 
for stars of all spectral types and all activity levels.  Variability in chromospheric activity
also contributes to some of the observed RMS scatter.}
\end{figure}
\clearpage

\begin{figure}
\plotone{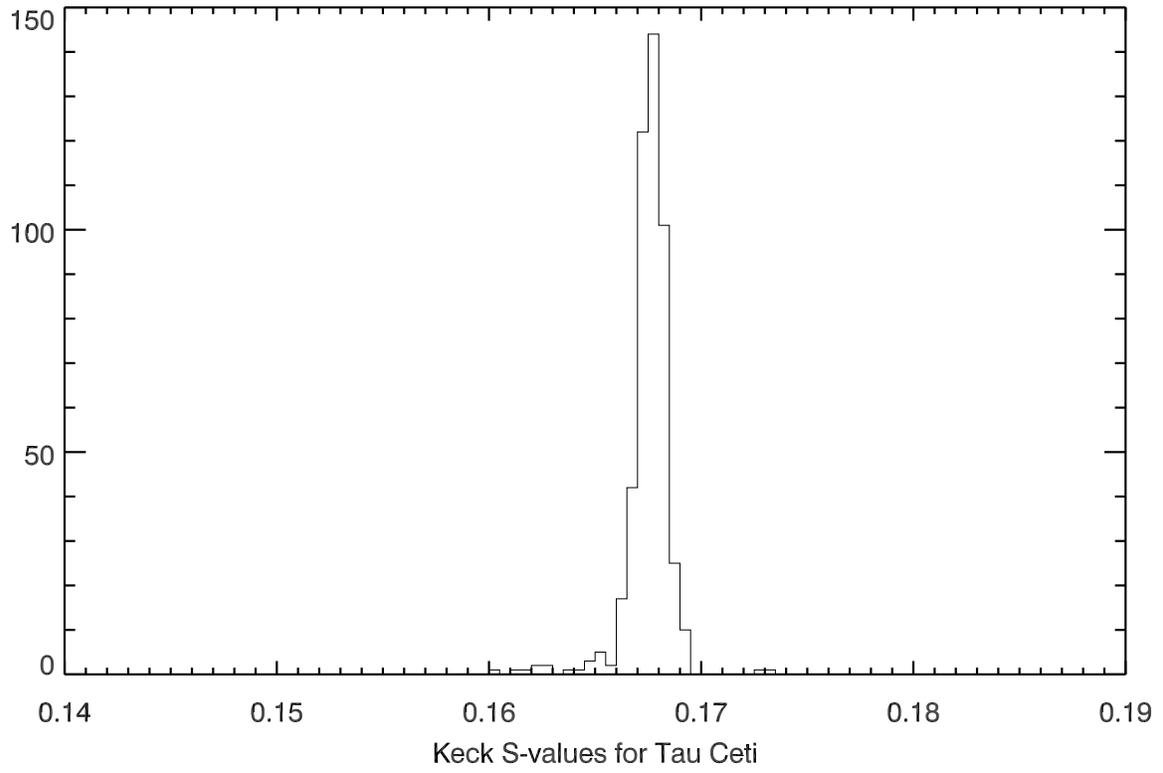}
\figcaption{A histogram of \shk\ for the chromospherically inactive star, $\tau$ Ceti.
The median \shk\ value from Keck is 0.167 and the FWHM of the distribution is roughly 
0.009 or 1\%.  This provides an empirical estimate for the errors assigned to activity measurements.}
\end{figure}
\clearpage

\begin{figure}
\plotone {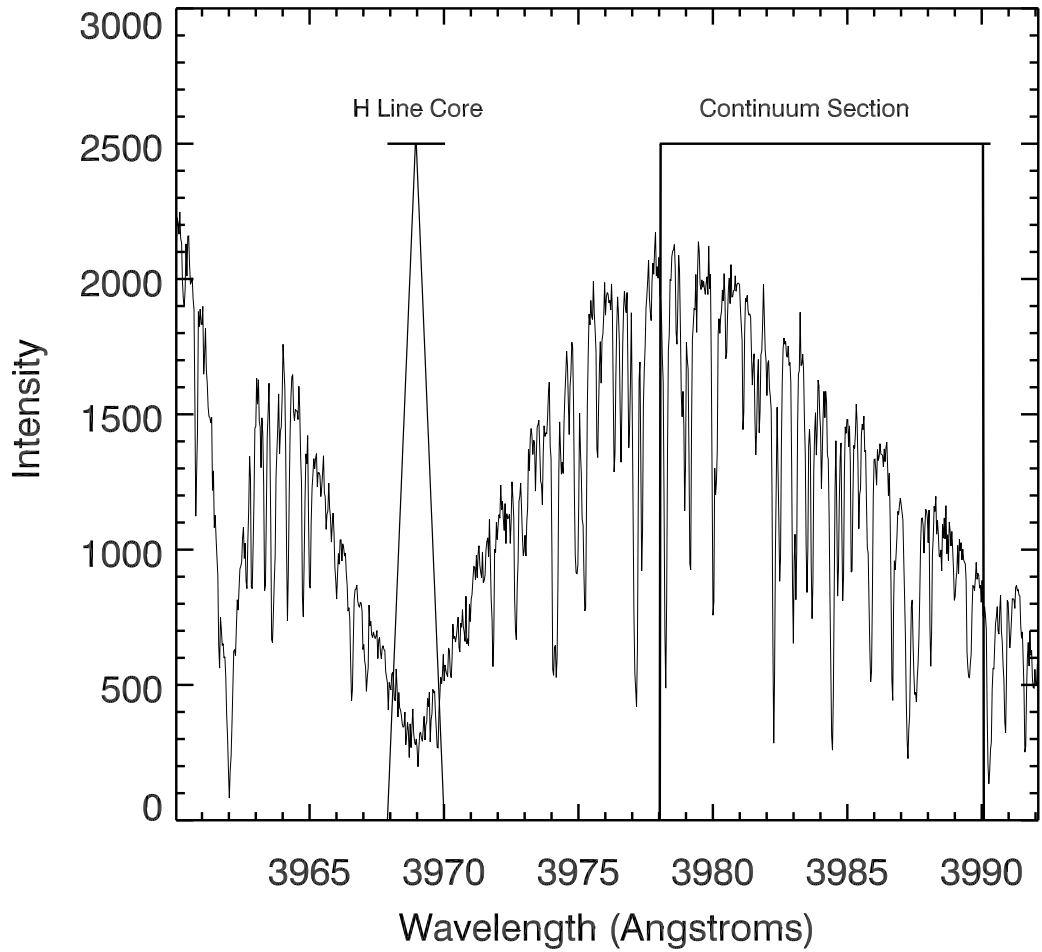}
\figcaption{Only one echelle order contains adequate signal for analysis of \shk\ at Lick Observatory. 
The ``continuum'' section as well as the weighting function for the H line core is indicated in this 
spectrum of $\tau$ Ceti. }
\end{figure}
\clearpage

\begin{figure}
\plotone{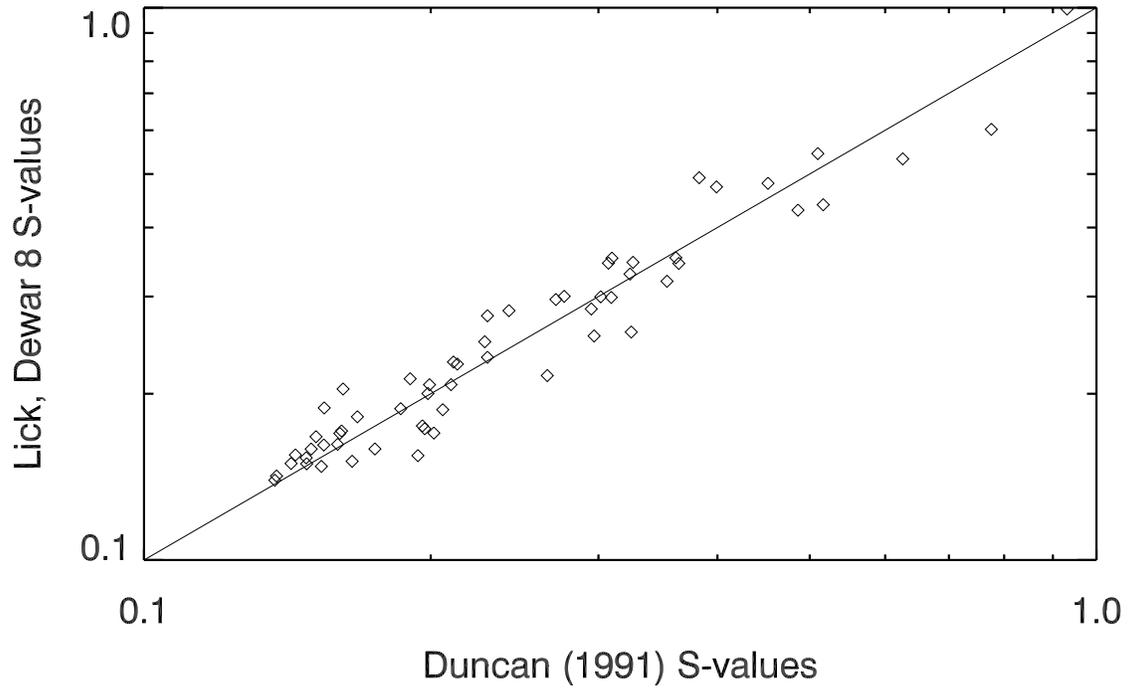}
\figcaption{The \sh\ values measured at Lick are based only on emission in the 
\caii\ H line. There were 83 stars in common between the Lick program and the Mt. Wilson 
H \& K survey that were used for the calibration. 
An RMS scatter of 11\% was found when fitting the coefficients for \sh\ at Lick. }
\end{figure}
\clearpage

\begin{figure}
\plotone{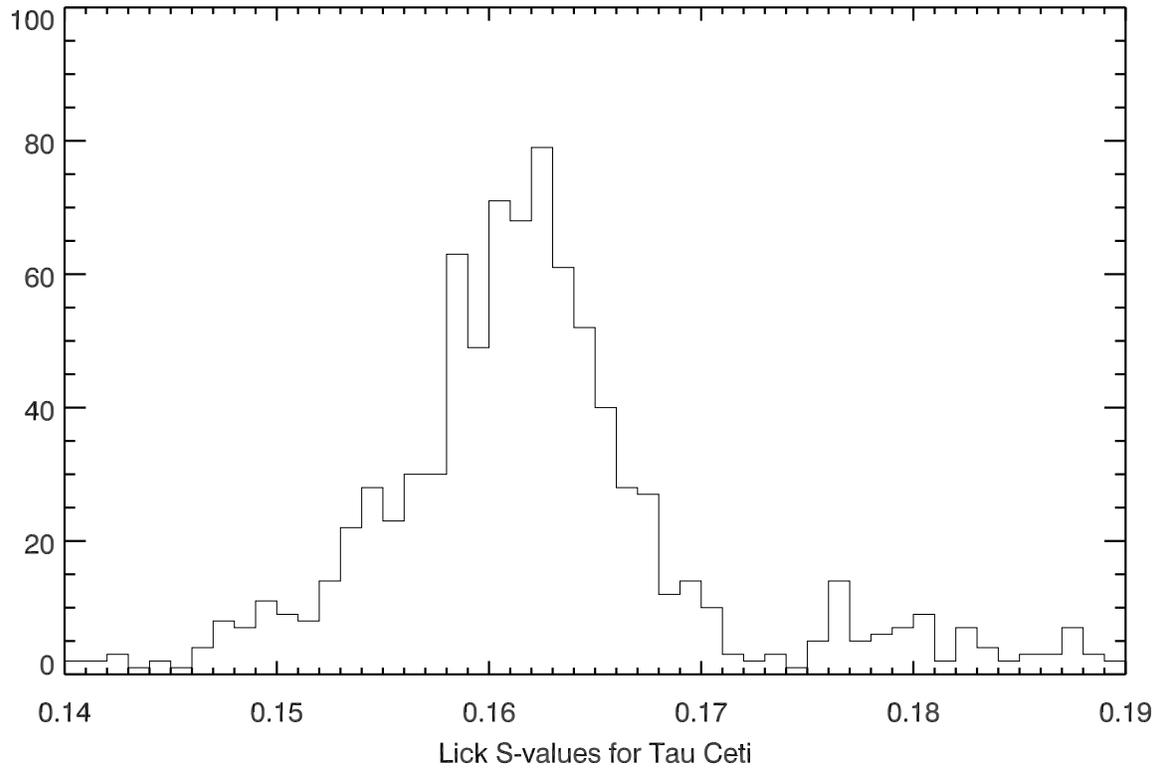}
\figcaption{The histogram of \sh\ values for $\tau$ Ceti from Lick Observatory has a median 
\sh\ value of 0.163. 
The FWHM of the measurement distribution is roughly 0.008, suggesting 
a single measurement precision of about 5\% for \sh.  }
\end{figure}
\clearpage

\begin{figure}
\plotone{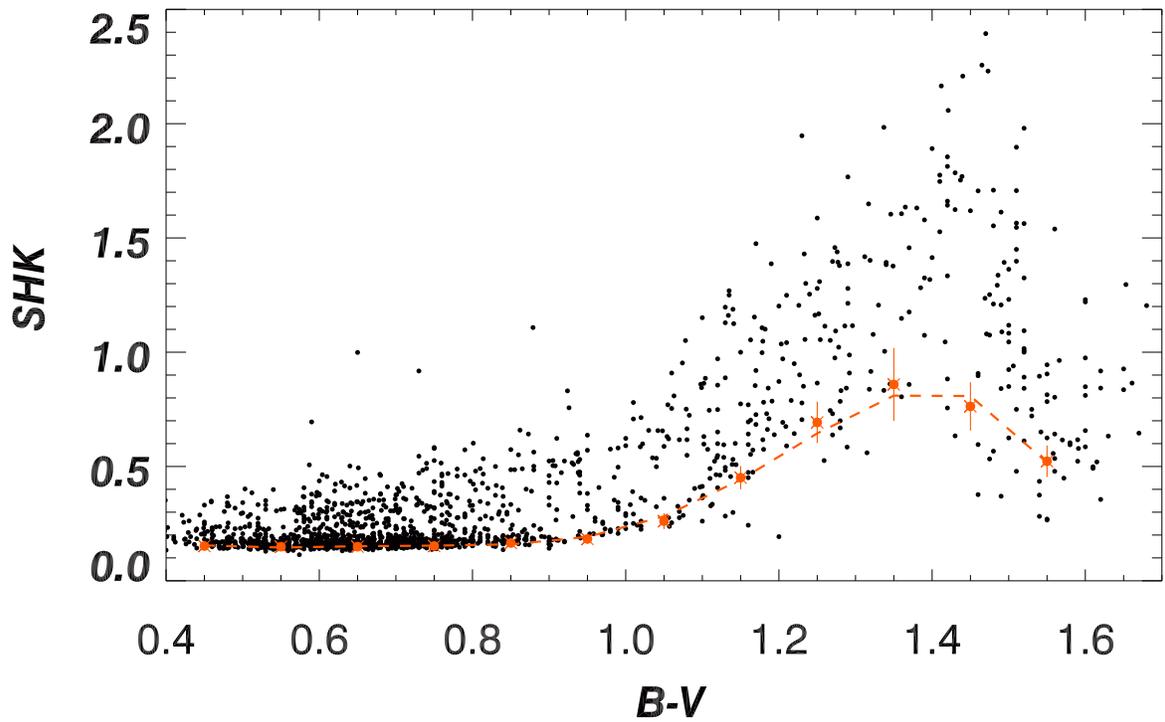}
\figcaption{The \shk\ values for main sequence stars are plotted as function of 
\bv\ color. The lower envelope of the \shk\ values is defined as the tenth percentile
in activity (red dashed line) and adopted as a basal activity level, \sten. There is a decrease
in \sten\ for stars redward of $\bv\ = 1.4$ that reflects a drop in chromospheric activity.}
\end{figure}
\clearpage 

\begin{figure}
\plotone{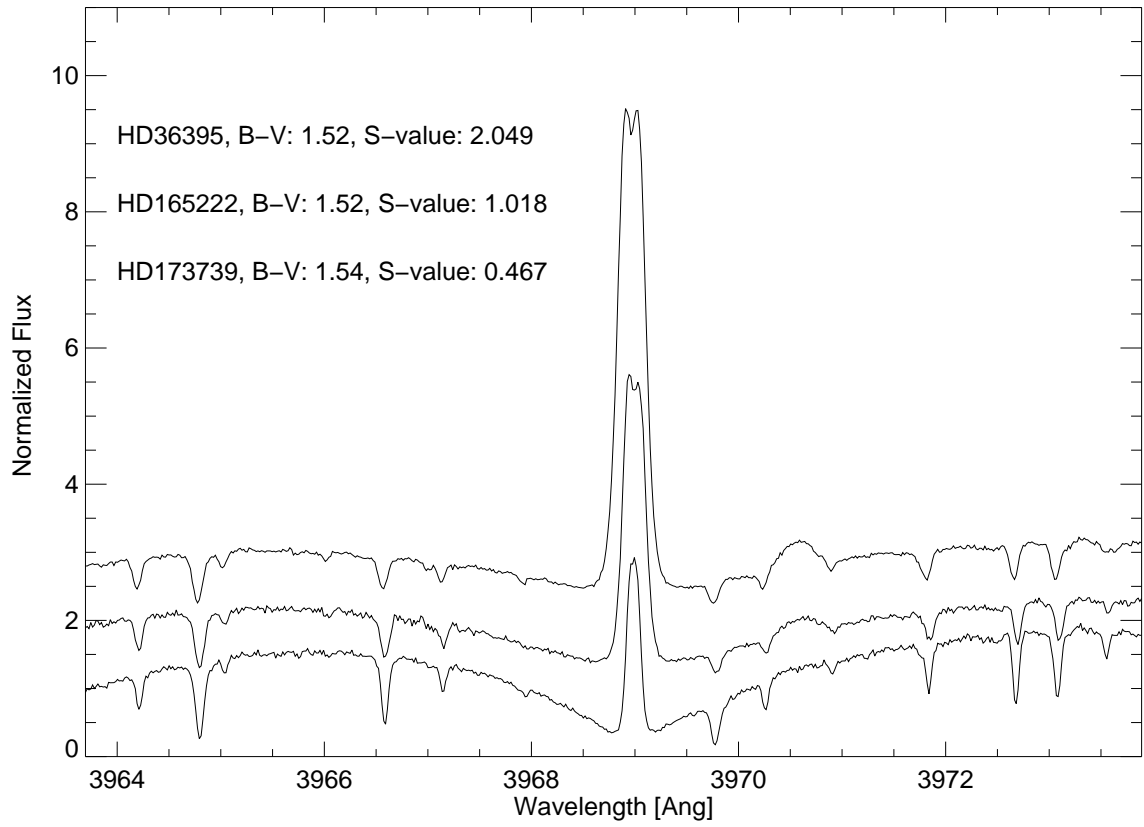}
\figcaption{Emission lines are shown for three stars with approximately the 
same \bv\ and different activity levels that demonstrate a range in chromospheric activity. }
\end{figure}
\clearpage 

\begin{figure}
\plotone{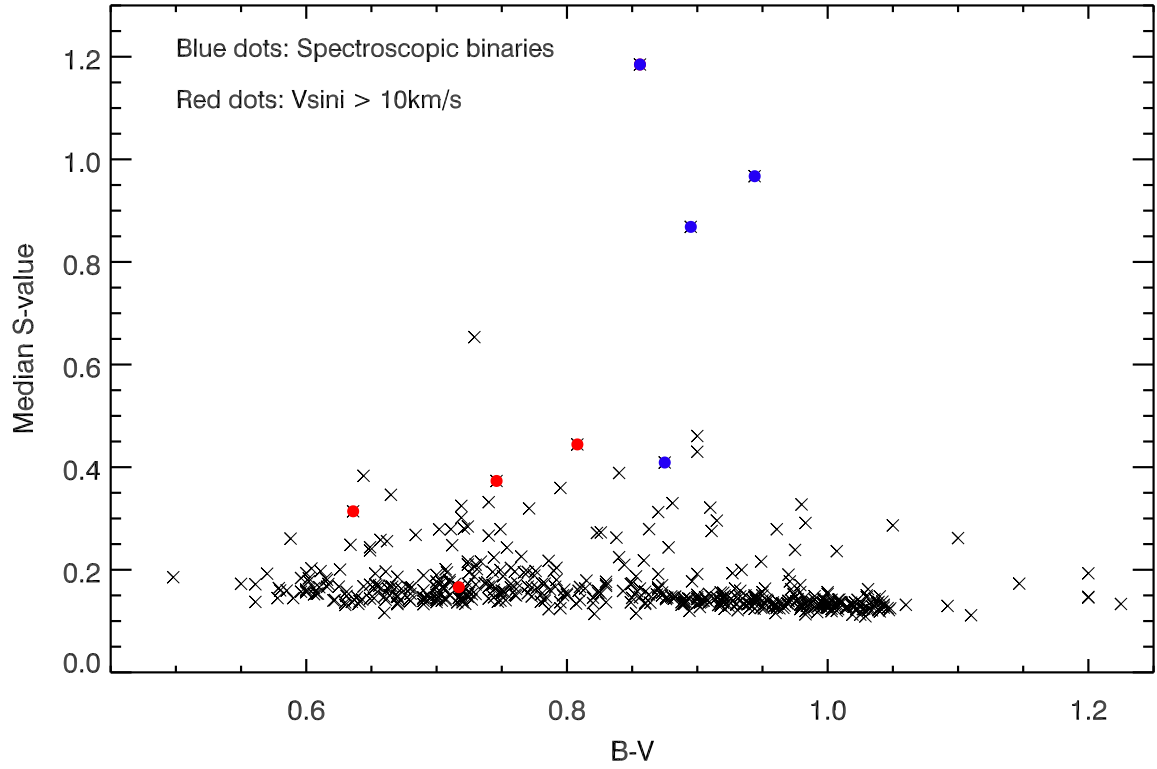}
\figcaption{The \shk\ values of the vast majority of subgiants is consistent with 
low chromospheric activity.  About ten percent of subgiants are active. Some of the 
active subgiants are rapid rotators or close binaries. }
\end{figure}
\clearpage 

\begin{figure}
\plotone{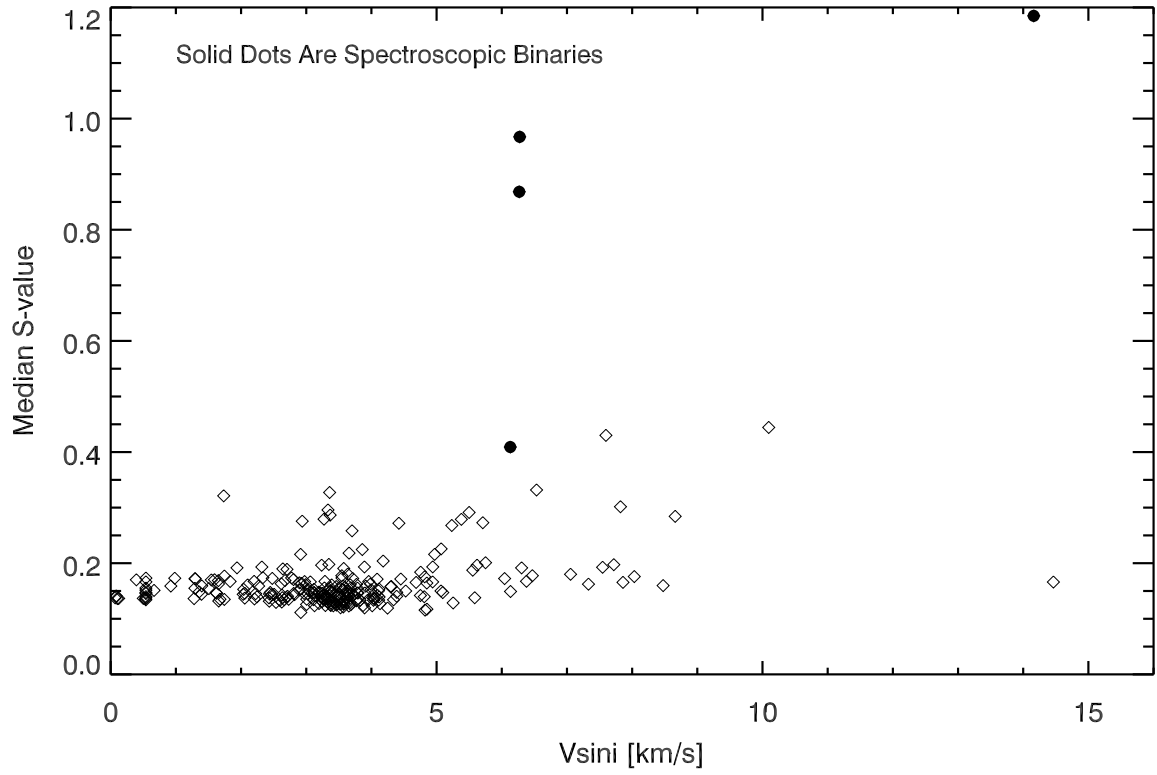}
\figcaption{The chromospheric activity in subgiants is plotted as a function of 
\vsini. Rotation and activity appear to be correlated. The presence of a close binary
stellar companion is also correlated with chromospheric activity.}
\end{figure}
\clearpage 

\begin{figure}
\plotone{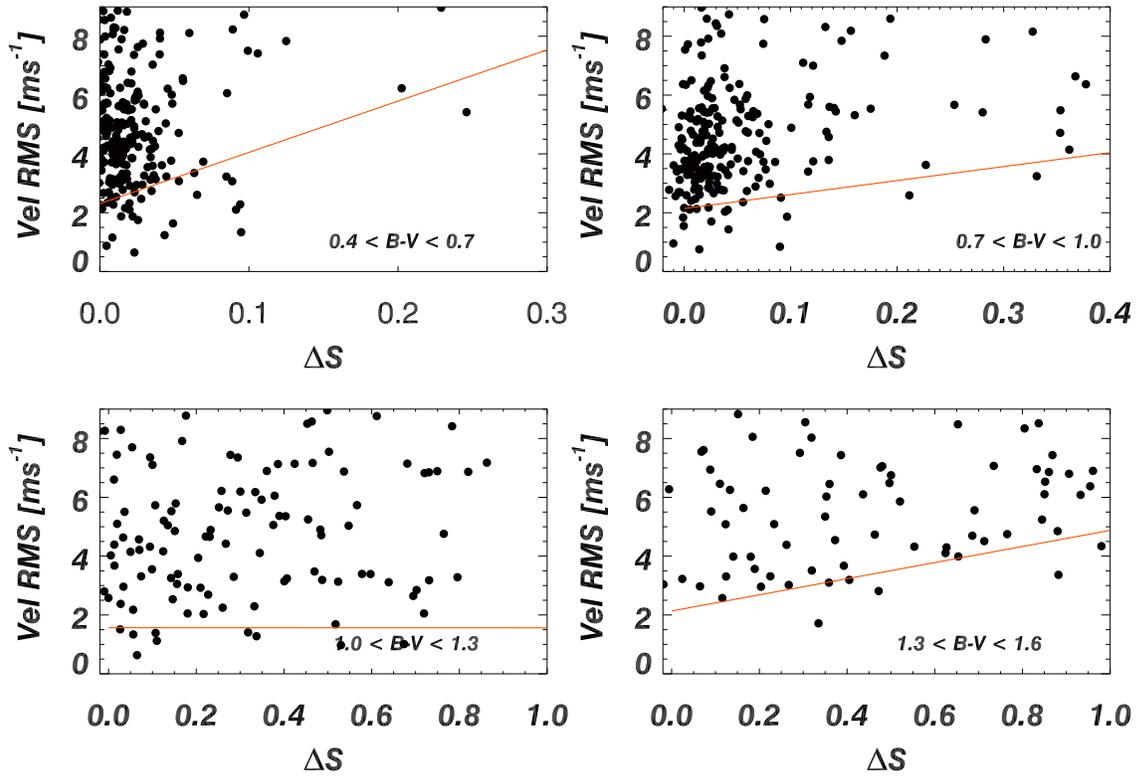}
\figcaption{Radial velocities are plotted for main sequence stars as a function of 
\sten. The left plot shows stars with $0.4 < \bv\ < 1.0$ and the right plot shows 
data for stars with $1.0 < \bv\ < 1.6$.  The floor of the velocity \rms\ is adopted 
as jitter for these stars (which may include instrumental systematic errors). 
Overall, the redder stars exhibit a lower level of jitter than blue stars. 
High \caii\ emission stars are rare among the bluer 
stars, but the velocity jitter is steeper for these stars. }
\end{figure}
\clearpage 

\begin{figure}
\plotone{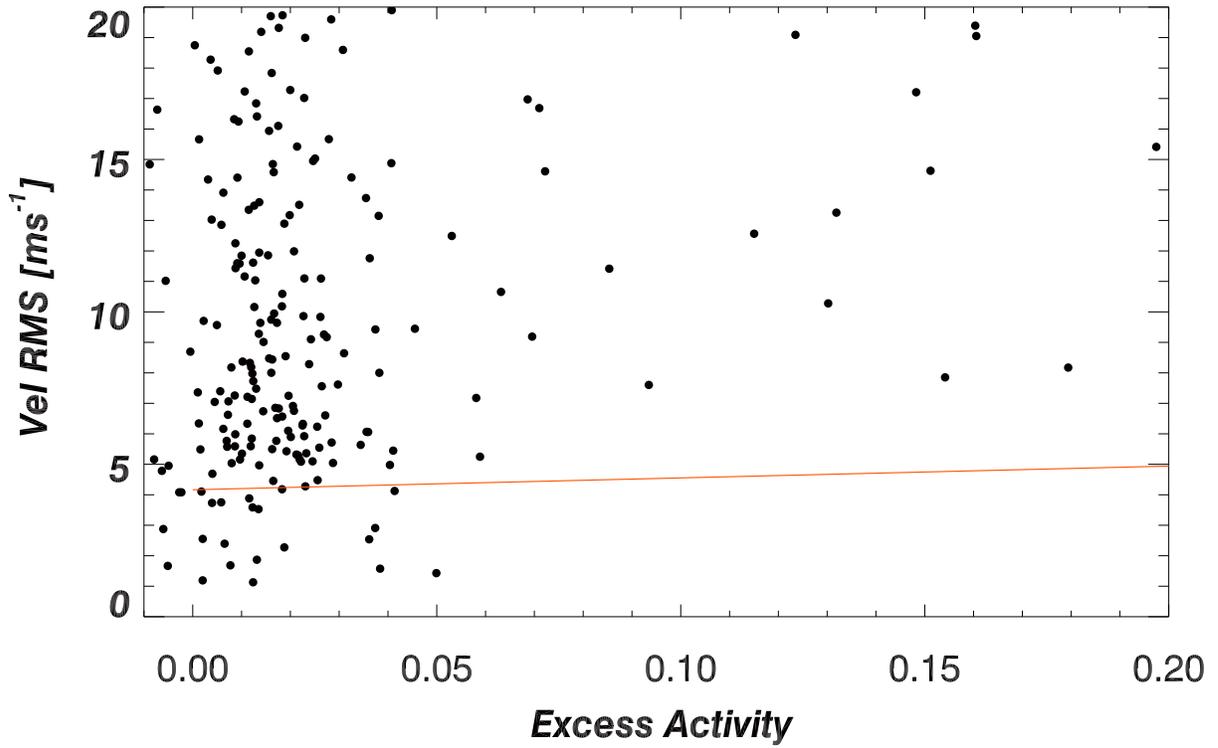}
\figcaption{Radial velocities are plotted for the \citet{j10} sample of subgiants.
Most subgiants are chromospherically inactive, so it is difficult to quantify the 
slope of jitter for these stars. However the subgiants appear to be only modestly 
more active than main sequence stars.  }
\end{figure}
\clearpage 

\begin{figure}
\plotone{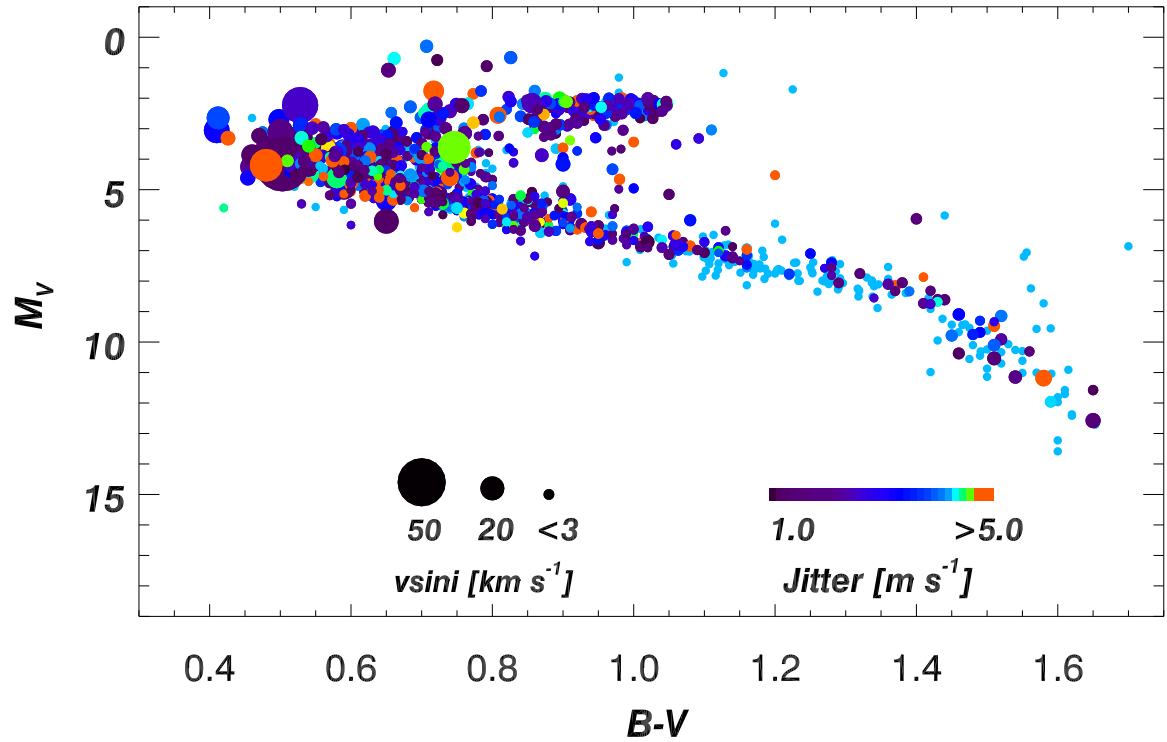}
\figcaption{The HR diagram for CPS stars.  The symbol color is scaled according to 
the jitter legend and symbol size is scaled to \vsini. The lower main sequence stars
have modest jitter, while FGK stars show a broader range in background noise. Most stars on the 
subgiant branch have low activity and low intrinsic jitter, however about ten percent of subgiants are 
chromospherically active.  }
\end{figure}
\clearpage 

\begin{figure}
\plotone{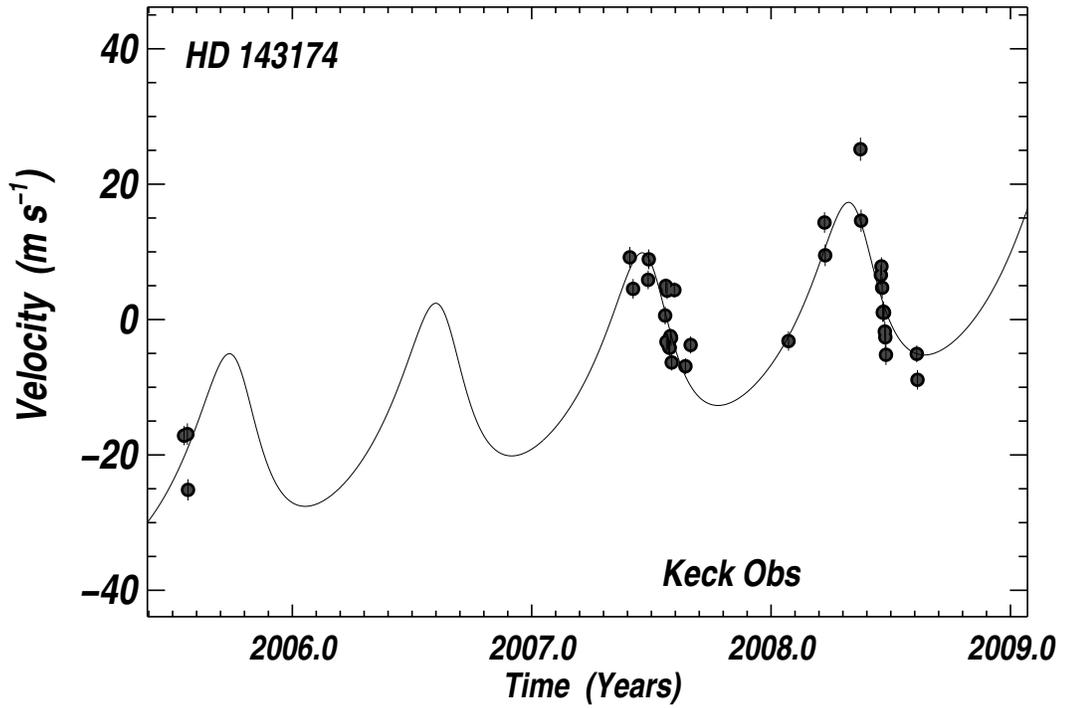}
\figcaption{Radial velocities for HD~143174 exhibited velocity \rms\ of almost 10 \ms\ and 
clear structure in the time series Doppler measurements that was initally well-fit by 
a Keplerian model. The activity level for this $\bv = 0.61$ star has a \dels\ = 0.093 
and expected jitter of just 3.9 \ms.}
\end{figure}
\clearpage

\begin{figure}
\plotone{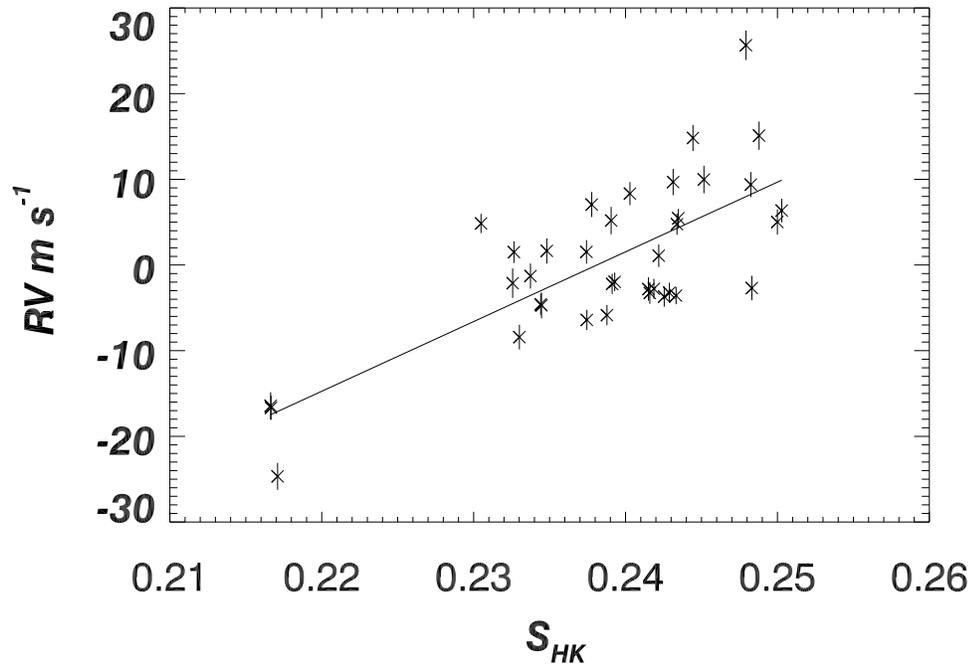}
\figcaption{HD~143174 shows a suspicious correlation between chromospheric activity index, \shk\ and 
radial velocities.  }
\end{figure}
\clearpage


\begin{deluxetable}{rrrrrrrrrr}
\tablecaption{ S-values and Derived Quantities\label{T1} }
\tablewidth{0pt}
\tablehead{
  \colhead{Star} &
  \colhead{ }  & 
  \colhead{ }  & 
  \colhead{Median}  & 
  \colhead{Jitter}  &
  \colhead{Median}  & 
  \colhead{ }  & 
  \colhead{$P_{rot}$} & 
  \colhead{ }  & 
  \colhead{ }  \\

  \colhead{name\tablenotemark{a}} &
  \colhead{HIP}  & 
  \colhead{B-V} & 
  \colhead{S} & 
  \colhead{$m s^{-1}$} &
  \colhead{\rhk} & 
  \colhead{$\Delta M_{v}$} &  
  \colhead{(days)} & 
  \colhead{log Age}  & 
  \colhead{Obsv} \\ 
}
\startdata
      GJ 26 &     ---  & 1.540  & 0.895  & 2.844  &      ---  &  0.17  &   ---  &   ---  & K \\
       38A  &     473  & 1.410  & 1.775  & 4.542  &      ---  &  1.01  &   ---  &   ---  & K \\
       38B  &     ---  & 1.410  & 1.747  & 4.475  &      ---  &  0.15  &   ---  &   ---  & K \\
      GJ 47 &     ---  & 1.390  & 1.074  & 2.885  &      ---  &  -1.4  &   ---  &   ---  & K \\
      GJ 48 &    4856  & 1.460  & 0.907  & 2.483  &      ---  &  -0.9  &   ---  &   ---  & K \\
      GJ 49 &    4872  & 1.500  & 2.947  & 7.510  &      ---  &  0.51  &   ---  &   ---  & K \\
    GJ 83.1 &     ---  & 1.820  & 14.72  & 0.000  &      ---  &  0.76  &   ---  &   ---  & K \\
      GJ 87 &   10279  & 1.430  & 0.634  & 2.100  &      ---  &  -0.8  &   ---  &   ---  & K \\
       105  &     490  & 0.600  & 0.380  & 6.581  &   -4.344  &  -0.0  &     3  &  0.19  & K \\
    GJ 105B &     ---  & 1.620  & 0.843  & 3.717  &      ---  &  -0.1  &   ---  &   ---  & K \\
    GJ 107B &     ---  & 0.930  & 1.766  & 9.987  &      ---  &  -3.0  &   ---  &   ---  & K \\
     GJ 109 &   12781  & 1.550  & 0.814  & 2.736  &      ---  &  -0.2  &   ---  &   ---  & K \\
       166  &     544  & 0.750  & 0.434  & 3.586  &   -4.384  &  0.00  &     7  &  0.26  & K \\
\enddata
\tablenotetext{a}{Star names are HD unless otherwise given. Values of B-V are from the 
Hipparcos Catalog. Derived quantities are only given for stars in the range 
$ -4.0 < \rhk\ < -5.1 $.  Obsv is K for Keck Observatory or L for Lick Observatory.}
\end{deluxetable}
\clearpage

\begin{deluxetable}{llcccrc}
\tablecaption{ S-values and Julian Date - 2440000 \label{T2}}
\tablewidth{0pt}
\tablehead{
  \colhead{Star} &
  \colhead{ } &
  \colhead{JD} & 
  \colhead{ } &
  \colhead{ } &
  \colhead{ } &
  \colhead{ } \\

  \colhead{name} &
  \colhead{HIP}  & 
  \colhead{- 2440000.} & 
  \colhead{S value} & 
  \colhead{\rhk } & 
  \colhead{SNR} &
  \colhead{Obsv} \\ 
}
\startdata
GJ 26  & ---  &  13238.978  &  0.686  &  ---  &  17  &   K  \\
GJ 26  & ---  &  13338.806  &  0.614  &  ---  &  13  &   K  \\
GJ 26  & ---  &  13369.738  &  0.661  &  ---  &  15  &   K  \\
GJ 26  & ---  &  13426.728  &  0.598  &  ---  &  8  &   K  \\
GJ 26  & ---  &  13723.735  &  1.045  &  ---  &  9  &   K  \\
GJ 26  & ---  &  13961.013  &  0.939  &  ---  &  14  &   K  \\
GJ 26  & ---  &  13981.934  &  0.939  &  ---  &  15  &   K  \\
GJ 26  & ---  &  13982.964  &  0.845  &  ---  &  14  &   K  \\
\enddata
\end{deluxetable}
\clearpage

\end{document}